\documentclass[useAMS,usenatbib,usegraphicx]{mn2e}

\usepackage{amsmath,graphicx,amssymb,natbib}
\usepackage{times}

\newcommand\cpstar{a~Cen}
\newcommand{\zav}[1]{\left(#1\right)}
\newcommand{\hzav}[1]{\left[#1\right]}
\newcommand\intvidpo{\!\!\int\limits_{\begin{array}{c}\text{\scriptsize
visible}\\[-2mm]\text{\scriptsize surface}\end{array}}\!\!}

\newlength\staretab
\newcommand\zepsilon{\{\varepsilon\}}

\makeatletter
\def\sgn{\mathop{\operator@font sgn}\nolimits}
\makeatother
\newcommand\apj{ApJ}
\newcommand\apjs{ApJS}

\DeclareMathAlphabet{\mathsc}{OT1}{cmr}{m}{sc}
\def\testbx{bx}%
\DeclareRobustCommand{\ion}[2]{%
\relax\ifmmode
\ifx\testbx\f@series
{\mathbf{#1\,\mathsc{#2}}}\else
{\mathrm{#1\,\mathsc{#2}}}\fi
\else\textup{#1\,{\mdseries\textsc{#2}}}%
\fi}

\title{Distorted surfaces of magnetic helium-peculiar stars: An
application to a~Cen}
\author[Krti\v cka et al.]{J.~Krti\v cka,$^1$ 
        Z.~Mikul\'a\v sek,$^1$ M.~Prv\'ak,$^1$ 
        E.~Niemczura,$^2$ 
        F.~Leone,$^{3,4}$ and G.~Wade$^5$\\
$^1$Department of Theoretical Physics and Astrophysics,
        Masaryk University,  CZ-611\,37 Brno, Czech Republic,\\
$^2$Astronomical Institute, Wroc\l{}aw University, 51-622 Wroc\l{}aw, Poland,\\
$^3$Università di Catania, Dipartimento di Fisica e Astronomia, Sezione
       Astrofisica, I-95123 Catania, Italy,\\
$^4$INAF - Osservatorio Astrofisico di Catania, I-95123 Catania, Italy,\\
$^5$Department of Physics and Space Science, Royal Military College of Canada,
        P.O. Box 17000, Station Forces, Kingston, Ontario K7K 7B4, Canada}
\begin{document}

\date{\today}

\pagerange{\pageref{firstpage}--\pageref{lastpage}} \pubyear{2011}

\maketitle

\label{firstpage}

\begin{abstract}

Helium-peculiar magnetic chemically peculiar stars show variations of helium
abundance across their surfaces. As a result of associated atmospheric scale
height variations, the stellar surface becomes distorted, with helium-rich
regions dented inwards. Effectively, on top of flux variations due to opacity
effects, the depressed helium-rich surface regions become less bright in the
optical regions and brighter in the ultraviolet. We study the observational
effects of the aspherical surface on the light curves of a~Cen. We simulate the
light curves of this star adopting surface distributions of He, N, O, Si, and Fe
derived from Doppler mapping and introducing the effect of distortion
proportional to helium abundance. We show that while most of the optical and UV
variations of this star result from flux redistribution due to the non-uniform
surface distributions of helium and iron, the reduction of light variations due
to the helium-related surface distortion leads to a better agreement between
simulated optical light curves and the light curves observed with the BRITE
satellites.

\end{abstract}

\begin{keywords}
stars: chemically peculiar -- stars: early type -- stars:
variables -- stars: individual \cpstar
\end{keywords}

\section{Introduction}

During their lifetimes, the surfaces of chemically peculiar stars on the upper
main sequence acquire significantly different chemical composition than the
circumstellar clouds from which they were born. This results from the competing
processes of radiative levitation and gravitational settling in their
(primarily) radiative atmospheres \citep{vadog,mirivi,ales}. These processes
separate radiatively-supported elements from those that are deposited in the
deeper stellar layers.

The chemical separation affects a fraction of the stellar surface, which varies
with the age of the star \citep{vimir}. The peculiarities can appear only in
layers which are sufficiently stable against mixing processes. As a consequence, 
chemically peculiar stars are typically slow rotators and mostly avoid close
binaries \citep{nosic}. Moreover, the surface layers of chemically peculiar
stars are frequently stabilized by magnetic fields that are likely of fossil origin
\citep{morbob,wamimes,grunmimes} and may also originate in stellar mergers
\citep{botu,schneimer}.

Chemical elements are not distributed evenly across the surface of chemically
peculiar stars. They predominantly appear in large surface spots (patches), and the
abundance contrast across the surface can approach
several orders of magnitude \citep{luftb,kocuvir}. This results in spectroscopic
line profile variability modulated by the stellar rotation. The associated flux
redistribution, which stems from the dependence of opacity on wavelength, leads
to periodic light variability at the level of a few percent \citep{mycuvir,prvalis}.

Despite the order-of-magnitude variation of abundances in the stellar
atmosphere, the structural changes associated with the peculiar composition are
generally only modest. The reason is that hydrogen remains the dominant element
throughout the stellar atmosphere for most stars. It is only in helium-rich
stars that the dominant hydrogen is replaced by helium to a significant extent
\citep{choch,mysigorie}. This leads to up to a factor-of-two increase of mean
molecular weight and hence a factor-of-two reduction of atmospheric scale height
\citep{norba,mihacen}. These are significant changes that can be most easily
detected in photometry.

The inclusion of such distorted stellar surfaces could be important for our
understanding of the light curves of chemically peculiar (CP) stars. According
to the current paradigm, the light variability results from the flux
redistribution from the far-UV to near-UV and visible domains due to the
bound-bound (lines, mainly iron or chromium) and bound-free (ionization, mainly
silicon and helium) transitions, horizontal surface distribution of elements,
and stellar rotation \citep{mycuvir,prvalis}. The simulated light curves of CP
stars derived from the integrated flux of model atmospheres computed for
abundances from Doppler maps typically agree with observations in the
ultraviolet (UV) and optical domains. However, there are some details that
remain unexplained, for example the difference between the Str\"omgren $u$
and $v$ light curves of $\theta$~Aur \citep{mythetaaur} or the out-of-eclipse
light variations of $\sigma$~Ori~E \citep{mysigorie}. In some of these cases a
distorted stellar surface may provide an explanation of the discrepancies.

Due to variation of surface helium abundance and the density scale height the
stellar surface becomes distorted. We study the effect of distorted surface on
the light curves of the helium-peculiar star \cpstar\ (V761~Cen, HR~5378,
HD~125823). The star \cpstar\ is of particular interest in this context because
it exhibits the most extreme variation of He abundance across its surface of any
known He-peculiar star \citep{bohl}. Hence it is the best-suited target to
identify the potential effects of surface distortion. 

\section{The shape of the stellar surface}

The distortion of the stellar surface is connected with the variation of density
scale height 
\begin{equation}
H=\frac{kT}{g\mu m_\text{H}},
\end{equation}
which is given by the surface gravity $g$, hydrogen mass $m_\text{H}$, and mean
molecular weight $\mu$. In a helium-dominated gas, the mean molecular weight
changes from about 0.6 for the solar chemical composition to about 1.3 for fully
ionized helium. This implies a decrease of the density scale height by a factor
of up to about 2 \citep{norba,mihacen}. However, the real change of the stellar
radius is expected to be much smaller, because helium dominates only a fraction
of the stellar envelope.

To quantitatively estimate the influence of variable molecular weight on the
shape of the stellar surface we used MESA evolutionary models
\citep{mesa1,mesa2} to calculate the interior structure of \cpstar. We adopted
the zero age main sequence mass $M_0=5.6\,M_\odot$ and let the model evolve for
32\,Myr. These values correspond to the parameters of \cpstar\ as determined by \citet{kobacu}.

We assume that the helium-rich region extends from the stellar surface up to the
depth where the gas energy density $3\rho kT/(2\mu m_\text{H})$ starts to
dominate the magnetic field energy density $B^2/(8\pi)$. One can expect that
below this region the mixing processes do not allow for strong helium
enrichment. The depth of this region was derived from the MESA model assuming
magnetic field strength of 8\,kG, which corresponds to the maximum modulus of
the surface field of \cpstar\ as derived from the magnetic Doppler imaging (MDI,
\citealt{wadebrite}, Huang et al., in preparation). In a particular stellar
model, this happens at the radius $0.987\,R_\ast$ corresponding to a Rosseland
optical depth $\tau_\text{Ross}=460$. We modified the derived MESA model by
replacing hydrogen and helium mass fraction in the layers dominated by magnetic
field by $X=0$ and $Y=0.98$, respectively, and we let the model evolve for an
arbitrarily short time of 1\,yr. This should be long enough to allow the model
to relax to a new hydrostatic structure of the outer layers, because the
timescale of the approach to hydrostatic equilibrium is order of days for the
whole star. The change of surface chemical composition resulted in a negligible
change of the stellar luminosity, but the stellar radius $R_\ast$ decreased by
about $0.84\%$ and the effective temperature increased by roughly $\Delta
T_\text{eff}=80\,$K.

\begin{figure}
\centering \resizebox{\hsize}{!}{\includegraphics{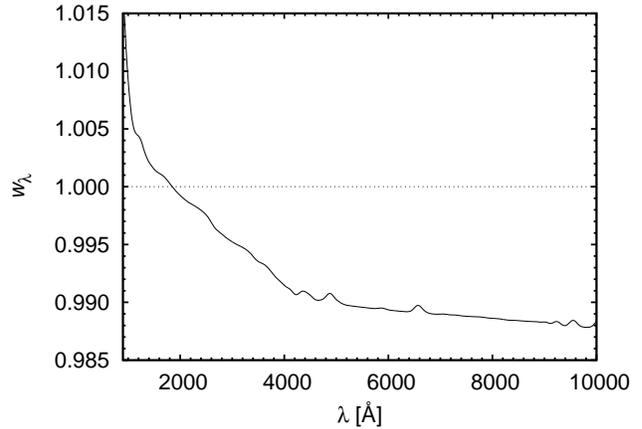}}
\caption{The ratio of emergent fluxes from the hotter (by 80\,K) depressed part
of the stellar surface and flux from the normal part of the surface corrected
for the different areas of these surfaces (Eq.~\eqref{warpeqdef}). The fluxes
were smoothed by a Gaussian filter with width 100\,\AA\ to show the changes in
continuum. The dotted line denotes a ratio of unity.}
\label{tokpodil}
\end{figure}

The total brightness variations due to the distorted surface is a combination of
the effect of the higher effective temperature and smaller radius, which
compensate each other to keep the luminosity constant. The change of brightness
as a function of wavelength can be described by a ratio of emergent fluxes
$H_\lambda$ calculated for the normal stellar effective temperature
$T_\text{eff}$ and for the effective temperature increased by $\Delta
T_\text{eff}$,
\begin{equation}
\label{warpeqdef}
w_\lambda=\zav{\frac{R_\text{w}}{R_\ast}}^2
\frac{H_\lambda(T_\text{eff}+\Delta T_\text{eff})}{H_\lambda(T_\text{eff})},
\end{equation}
where the geometrical factor accounts for a smaller radius $R_\text{w}$ at the
location of the distortion. From Fig.~\ref{tokpodil} it follows that in the
far-ultraviolet regions the temperature effect dominates due to the
redistribution; therefore these wavelength regions become brighter. On the other
hand, in the visible and near-ultraviolet regions the geometric factor dominates
and these layers become fainter. These changes contribute to the larger flux
variations caused by the opacity effects \citep{myhd37776}, but affect only
helium-rich regions of the stellar surface.

The distortion of the stellar surface contributes to the deviation from the
spherical symmetry. Because the surface layers are rigidly rotating due to the
magnetic field, the surface distortion leads to the precession of the star. Such
problems were studied in the context of the influence of magnetic field on the
stellar structure \citep{meta,lajprec}. The timescale of the evolution of the
direction of the rotational axis is roughly given by the angular frequency of
precession $(2\pi/P)\,(I_\text{w}-I_0)/I_0$, where $P$ is the period of rotation
and $I_\text{w}$ and $I_0$ are moments of inertia about the normal to the
helium-rich region and about the principal rotation axis, respectively (assuming
helium spot at the equator). From this follows the typical timescale of about
$10^4\,$yr, which was determined using the moments of inertia calculated from
the output of the MESA models discussed above.

The precessional motion of the rotational axis can be observationally tested by
study of rotational period variations, but corresponding changes are very small
\citep{brzda}. On the other hand, the bulk of the star is not confined by the
magnetic field and the precessional motion is accompanied by interior flows,
which are oscillatory on the precession timescale \citep{meta,lajprec}. This
could possibly explain long timescales of rotational period evolution 
observed in some helium peculiar stars \citep{brzda,matbrzd}, that cannot be
explained by rotational braking by magnetized wind \citep{brzdud}. 

The stability considerations require that the star rotates along the axis with
largest corresponding moment of inertia, that is, with helium-rich regions
located around the poles. This does not appear in real stars implying either
long relaxation processes or the existence of another stronger surface
distortion.

The distortion of the stellar surface due to helium abundance variations may be
modified by the effect of the magnetic field \citep{mihacen}. While the poloidal
field likely does not significantly contribute to the pressure gradient due to
small variations of the poloidal field component in the atmosphere, the toroidal
field component may have more significant effect.

\section{Ultraviolet and visual variability}

\begin{table}
\caption{Parameters of a Cen (Huang et al., in preparation).}
\label{hvezda}
\begin{center}
\begin{tabular}{lc}
\hline
Effective temperature ${{T}_\mathrm{eff}}$ & ${19\,000}$\,K \\
Surface gravity ${\log g}$ (cgs) & ${4.0}$ \\
Inclination ${i}$ & ${70^\circ}$ \\
Microturbulent velocity & $2\,\text{km}\,\text{s}^{-1}$\\
Helium abundance&$-2.2<\varepsilon_\text{He}<1.5$ \\
Nitrogen abundance& $-4.3<\varepsilon_\text{N}<-2.9$ \\
Oxygen abundance& $-3.4<\varepsilon_\text{O}<-2.1$ \\
Silicon abundance& $-5.4<\varepsilon_\text{Si}<-3.0$ \\
Iron abundance&$-5.6<\varepsilon_\text{Fe}<-1.9$  \\
\hline
\end{tabular}
\end{center}
\end{table}

\subsection{The observed rotational variability of a Cen}

The magnetic helium weak/strong chemically peculiar star a Cen is known as
spectroscopic, photometric, and magnetic variable with a period of 8.817\,d
which was identified with its rotational period \citep{mihacen,catleo}. The
simultaneous analysis of 99\,497 relevant observational data covering the time
interval 1967--2016 shows that the observed variations can be satisfactorily
well approximated by the linear ephemeris
\begin{equation}
\text{JD}= 2\,451\,977.386(19) + 8.816\,966(10)\times E.
\end{equation}
The period analysis, which is based on $uvby$ \citep{noracen,peto,catleo},
Hipparcos \citep{esa97}, SMEI (Pigulski, priv.~comm.), and BRITE photometry, He
line strengths \citep{noracen,underacen,peto}, magnetic field measurements
\citep{borhel}, and our own He line spectroscopy and magnetic field
measurements, will be published elsewhere (Mikul\'a\v sek et al., in
preparation). The zero phase corresponds to the maximum strength of He lines. We
used this ephemeris for the simulations that follow\footnote{The rotational
period, derived from the above mentioned observations and new high-precision
data from TESS satellite, $P = 8.816\,991 (9)$\,d (for full period analysis see
Mikul\'a\v sek et al., in preparation) does not contradict the ephemeris used in
this paper.}.

\subsection{Simulation of the SED variability}

The simulation of the variability of the spectral energy distribution (SED) of
a~Cen is based on the model atmosphere code TLUSTY200 \citep{tlusty,lahub}. The
list of included elements, ionization states, and the atomic data used for the
atmosphere modelling are the same as used by \citet{bstar2006}. The data are
appropriate for B-type stars and they were mostly calculated within the Opacity
and Iron Projects \citep{topt,zel0}. Our calculations assumed the (fixed)
effective temperature and surface gravity given in Table~\ref{hvezda}. We
calculated a grid of LTE model atmospheres parametrized by abundances of helium,
nitrogen, oxygen, silicon, and iron (see Table~\ref{esit}). The abundances cover
the range of abundances found on the surface of a~Cen from the MDI, except for
the lowest abundances of helium, nitrogen, and oxygen. Our test showed that
these elements at their lowest abundances found from the MDI do not
significantly influence the emergent flux. For all other elements not included
in the MDI we assumed solar abundances after \citet{asp09}. In the model
atmospheres, we assumed the chemical stratification to be vertically
homogeneous, although vertical abundance gradients may exist in helium rich
stars \citep{vadog,lela,lecos}. In the following $\varepsilon_\text{el}$ denotes
abundances relative to hydrogen by number, i.e.,
$\varepsilon_\text{el}=\log\zav{N_\text{el}/N_\text{H}}$.

Synthetic spectra were calculated from the model atmospheres using the SYNSPEC45
code assuming the same abundances as the model atmosphere calculations. For the
grid of abundances specified in Table~\ref{esit}, we derived the
abundance-dependent specific intensities $I(\lambda, \theta, \zepsilon$) for
$20$ equidistantly-spaced values of $\cos\theta$, where $\theta$ is the angle
between the normal to the surface and the line of sight, and $\zepsilon$ denotes
vector of abundances, $\zepsilon= (\varepsilon_\text{He}, \varepsilon_\text{N},
\varepsilon_\text{O}, \varepsilon_\text{Si}, \varepsilon_\text{Fe})$. The
original linelist provided with SYNSPEC45 was extended by about 23 million iron
lines derived from theoretical and observational line lists downloaded in 2013
from the Kurucz website\footnote{http://kurucz.harvard.edu}.

\begin{table}
\caption{Individual abundances $\varepsilon_\text{He}$, $\varepsilon_\text{N}$,
$\varepsilon_\text{O}$, $\varepsilon_\text{Si}$, and $\varepsilon_\text{Fe}$ of
the model grid.}
\label{esit}
\begin{center}
\begin{tabular}{lr@{\hspace{2.5mm}}r@{\hspace{2.5mm}}r@{\hspace{2.5mm}}r@{\hspace{2.5mm}}r@{\hspace{2.5mm}}r@{\hspace{2.5mm}}r@{\hspace{2.5mm}}r@{\hspace{2.5mm}}r}
\hline
He& $-1.5$ & $-1.0$& $-0.5$& $0.0$& $0.5$& $1.0$ & $1.5$\\
N& $-3.9$ & $-3.4$ & $-2.9$\\
O& $-2.6$ & $-2.1$ \\
Si& $-5.5$ & $-5.0$ & $-4.5$ & $-4.0$ & $-3.5$ & $-3.0$ \\
Fe & $-5.9$ & $-5.4$ & $-4.9$ & $-4.4$ & $-3.9$ & $-3.4$ & $-2.9$ & $-2.4$ & $-1.9$\\
\hline
\end{tabular}
\end{center}
\end{table}

The basic photometric quantity observed at a distance $D$ from the star is the
radiative flux in a band $c$ \citep{hubenymihalas}
\begin{equation}
\label{vyptok}
f_c=\zav{\frac{R_*}{D}}^2\intvidpo I_c(\theta,\Omega)\cos\theta\,\text{d}\Omega,
\end{equation}
where $R_*$ is the stellar radius. The specific intensity $I_c(\theta,\Omega)$
in the band $c$ varies across the stellar surface. The intensity is obtained by
interpolating between intensities from the grid of $I_c(\theta, \zepsilon)$ at
each surface point with spherical coordinates $\Omega$. The intensities from the
grid are calculated as
\begin{equation}
\label{barint}
I_c(\theta,\zepsilon)= \int_0^{\infty}\Phi_c(\lambda) \,
I(\lambda,\theta,\zepsilon)\, \text{d}\lambda.
\end{equation}
The response function $\Phi_c(\lambda)$ for individual bands is derived by
either fitting the tabulated response functions for BRITE bands \citep{brite} or
simply assuming a Gaussian function for UV and $uvby$ variations (see
\citealt{myhd37776} for adopted coefficients for Str\"omgren bands).

For a comparison with the observed light variations we calculate the magnitude
difference in a given band defined as
\begin{equation}
\label{velik}
\Delta m_{c}=-2.5\,\log\,\zav{\frac{{f_c}}{f_c^\mathrm{ref}}},
\end{equation}
where $f_c$ is calculated from Eq.~\ref{vyptok}. Here
${f_c^\mathrm{ref}}$ is the reference flux obtained under the
condition that the mean magnitude difference over the rotational period is zero.

The influence of the aspherical distorted stellar surface was approximated by a function
\begin{equation}
\label{warpeq}
w_\lambda(\varepsilon_\text{He})=1+\hzav{\frac{1}{\pi}
\arctan\zav{\frac{\varepsilon_\text{He}}{\Delta\varepsilon}}+
\frac{1}{2}}\zav{w_\lambda^0-1},
\end{equation}
where we selected $\Delta\varepsilon=0.2$ and the values of $w_\lambda^0$ were
taken from Fig.~\ref{tokpodil}. The function was selected for convenience,
because it gives $w_\lambda=1$ for small $\varepsilon_\text{He}$ and
$w_\lambda=w_\lambda^0$ for large $\varepsilon_\text{He}$. The effect of
distortion was included in our simulations by multiplication of intensities
expressed by Eq.~\eqref{barint} by $w_\lambda$.

There is an additional effect connected with distortion of the stellar surface
in addition to the change of the surface brightness. As a result of distortion,
the normal to the surface no longer has a radial direction, but is tilted with
respect to the radial direction. This has to be accounted for when interpolating
between the specific intensities as a function of $\theta$ in
Eq.~\eqref{vyptok}. We included this effect in our models and it turns out that
this leads to a change of the light curve on the order of $10^{-4}\,\text{mag}$.
Because we do not know the detailed structure of the surface anyway, we
neglected this effect in the simulated light curves presented here.

\subsection{Influence of abundances on model atmospheres and emergent flux: the
case without surface asphericity}
\label{kaptoky}

\begin{figure}
\centering \resizebox{\hsize}{!}{\includegraphics{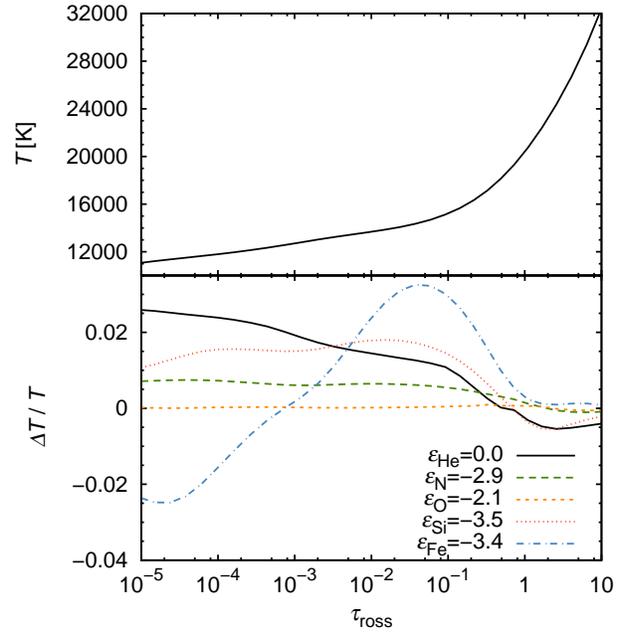}}
\caption{{\em Upper plot:} Dependence of temperature on the Rosseland
optical depth $\tau_\text{ross}$ in a reference model atmosphere with nearly
solar chemical composition ($\varepsilon_\text{He}=-1.0$,
$\varepsilon_\text{N}=-3.9$, $\varepsilon_\text{O}=-2.6$,
$\varepsilon_\text{Si}=-4.5$, and $\varepsilon_\text{Fe}=-4.4$). {\em Lower
plot}: The relative difference between the temperature in the model atmosphere
with modified abundance of a given element and the temperature in the reference
model.}
\label{tep}
\end{figure}

Individual elements influence the temperature distribution via light
absorption due to bound-bound (line) and bound-free (continuum) transitions. As
a result, the temperature in the continuum-forming region of the model
atmosphere (for $\tau_\text{ross}\approx0.1-1$) increases with increasing
abundance of given element (see Fig.~\ref{tep}). The influence of helium,
silicon, and iron is the strongest, while the effect of nitrogen and oxygen is
only marginal.


\begin{figure*}
\centering \resizebox{\hsize}{!}{\includegraphics{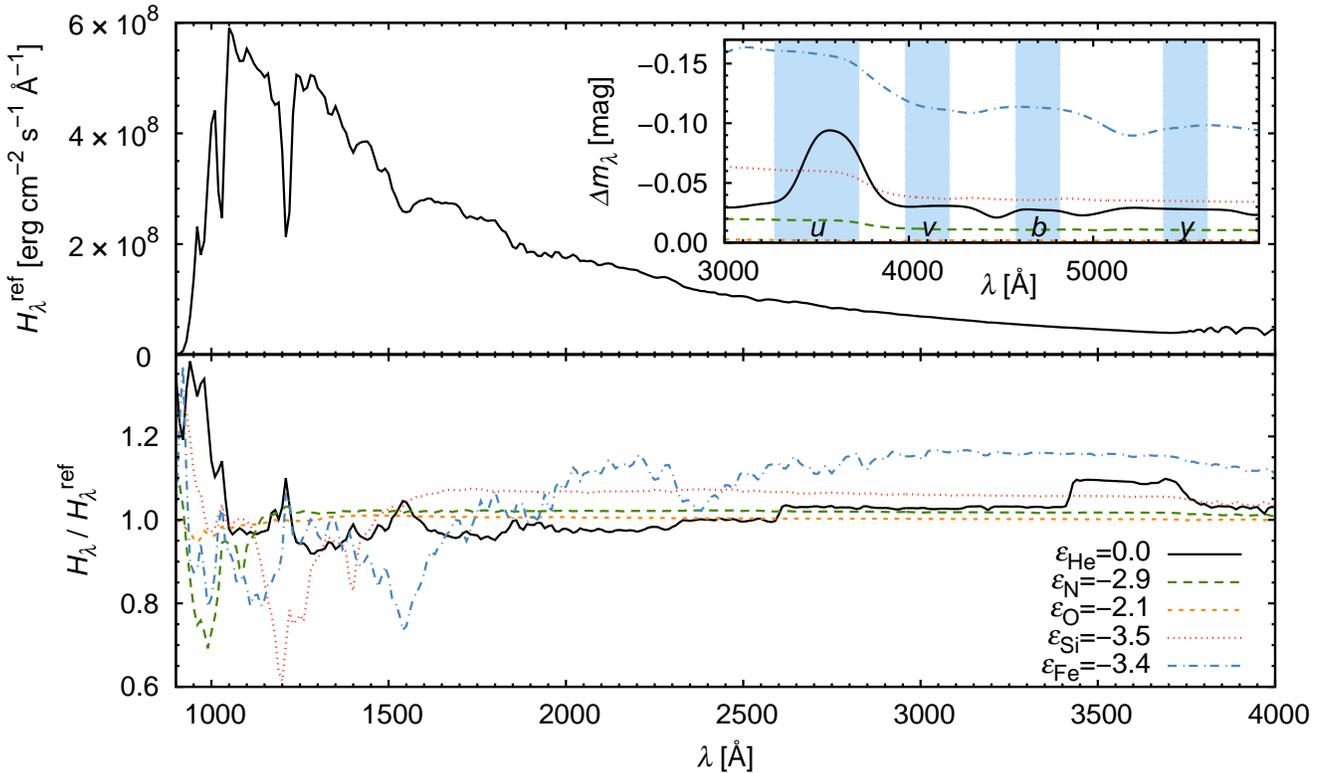}}
\caption{{\em Upper plot:} Emergent flux from a reference model atmosphere with
nearly solar chemical composition ($\varepsilon_\text{He}=-1.0$,
$\varepsilon_\text{N}=-3.9$, $\varepsilon_\text{O}=-2.6$,
$\varepsilon_\text{Si}=-4.5$, and $\varepsilon_\text{Fe}=-4.4$) as a function of
wavelength. {\em Lower plot}: The emergent flux from the model atmospheres with
enhanced abundances of individual elements relative to the flux from a reference
model. Fluxes were smoothed by a Gaussian filter with a dispersion of $10\,$\AA.
The inset in the upper plot shows magnitude difference (Eq.~\eqref{tokmagroz})
between the emergent fluxes calculated with enhanced abundances of individual
elements and the reference flux with near-solar chemical composition. The fluxes
were smoothed by a Gaussian filter with a dispersion of $100\,$\AA. }
\label{prvtoky}
\end{figure*}

Enhanced opacity in atmospheres with overabundance of individual elements
together with the associated modified temperature structure lead to the redistribution of
flux from wavelength regions with strong opacity of a given element to
regions with low opacity. Because the regions with strong opacity appear
typically in the far-UV region, the flux is typically redistributed from the far-UV
to the near-UV and optical regions (see Fig.~\ref{prvtoky}). Consequently,
overabundant spots are typically bright in the visual and near-UV bands, and are
dark in the far-UV bands. However, these variations are not monotonic due to
the non-monotonic dependence of opacity on wavelength.

These flux changes can be detected as SED variability in the UV and
optical regions. To understand the optical variations, we also plot the
relative magnitude difference
\begin{equation}
\label{tokmagroz}
\Delta m_\lambda=-2.5\log\zav{\frac{H_\lambda(\zepsilon)}
{H_\lambda^\text{ref}}},
\end{equation}
as a function of wavelength in Fig.~\ref{prvtoky}. Here $H_\lambda^\text{ref}$
is the reference flux calculated for nearly solar chemical composition
($\varepsilon_\text{He}=-1.0$, $\varepsilon_\text{N}=-3.9$,
$\varepsilon_\text{O}=-2.6$, $\varepsilon_\text{Si}=-4.5$, and
$\varepsilon_\text{Fe}=-4.4$). From Fig.~\ref{prvtoky} it follows that the
absolute value of the relative magnitude difference is the largest in the
$u$-band of the Str\"omgren photometric system due to helium and iron, while the
relative magnitude difference is nearly the same in the visual bands ($v$, $b$,
and $y$). An apparent maximum of the relative magnitude difference in the
$u$-band in the model with enhanced helium abundance is caused by the filling of
the Balmer jump. In the helium-rich models (typically for
$\varepsilon_\text{He}>0.5$), the hydrogen Balmer jump vanishes, while the jumps
due to helium become visible. From Fig.~\ref{prvtoky} it also follows that
nitrogen and oxygen do not significantly influence the visible light curve.

\subsection{Predicted visual and ultraviolet variations}

\begin{figure}
\centering
\resizebox{\hsize}{!}{\includegraphics{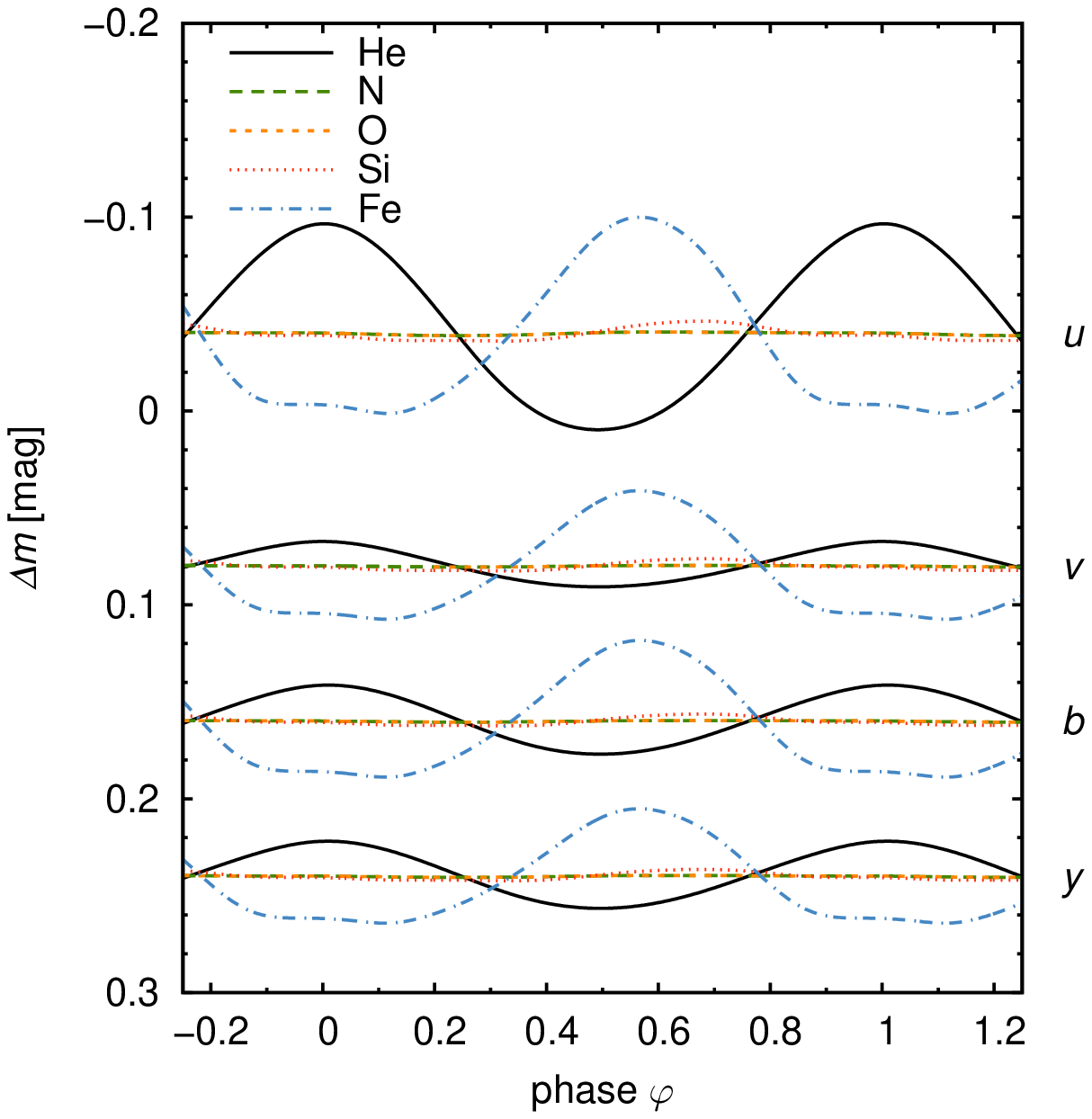}}
\caption{Simulated light variations of \cpstar\ in the Str\"omgren photometric
system calculated from abundance maps of one element only, neglecting the aspherical
surface. The abundance of other elements was fixed.   
Light curves in individual filters have been vertically
shifted to more clearly illustrate the light variability.}
\label{prv_hvvel}
\end{figure}

\begin{figure}
\centering
\resizebox{\hsize}{!}{\includegraphics{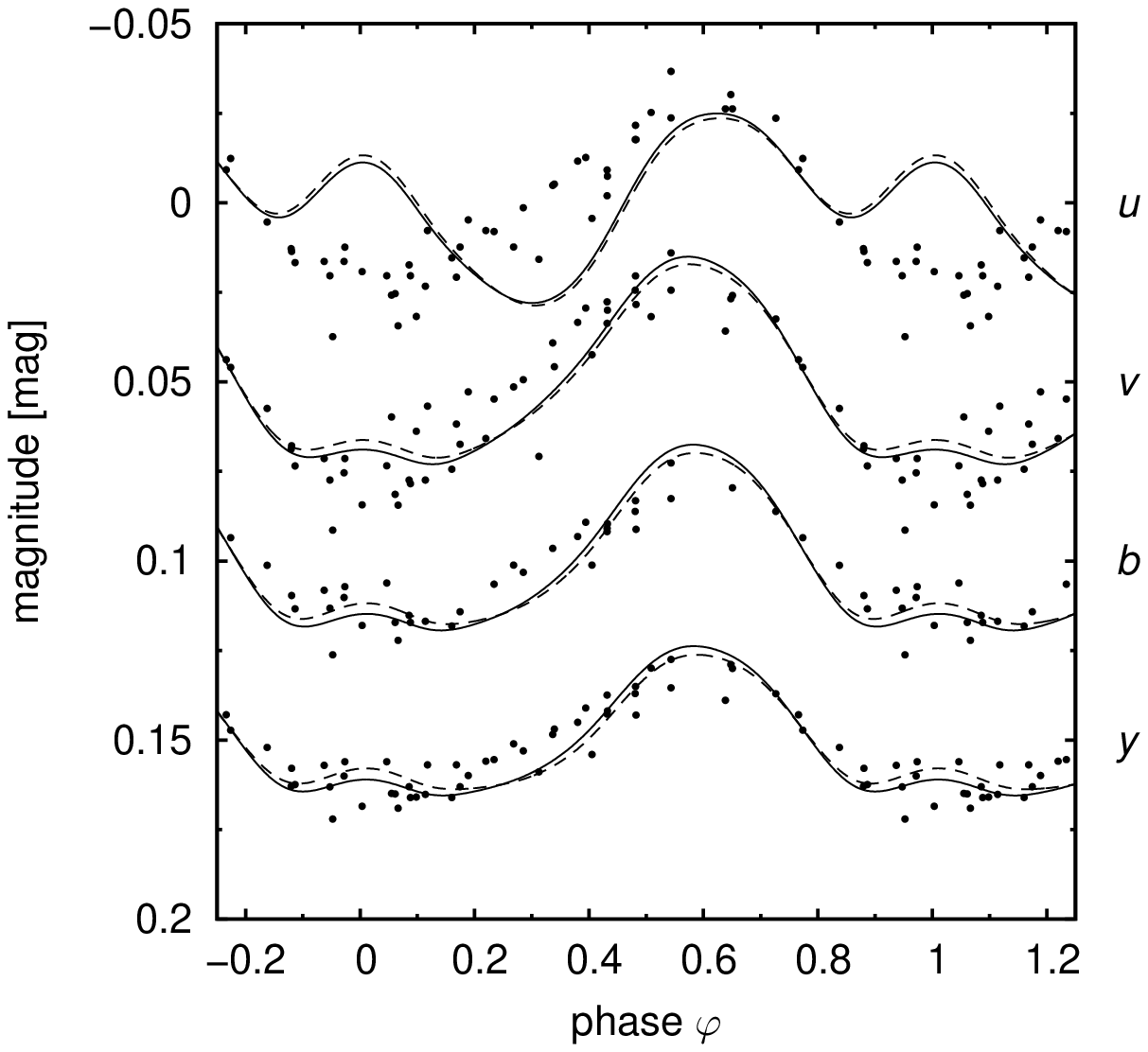}}
\caption{Predicted light variations of \cpstar\ with inclusion of surface
distortion (solid lines) and without distortion (dashed lines) computed taking
into account helium, nitrogen, oxygen, silicon, and iron surface abundance
distributions in comparison with observed light variations in the colors of the
Str\"omgren photometric system \citep{peto,catleo}. Light curves in
individual filters were vertically shifted to more clearly illustrate the light
variability.}
\label{acen_hvvel}
\end{figure}

\begin{figure}
\centering
\resizebox{\hsize}{!}{\includegraphics{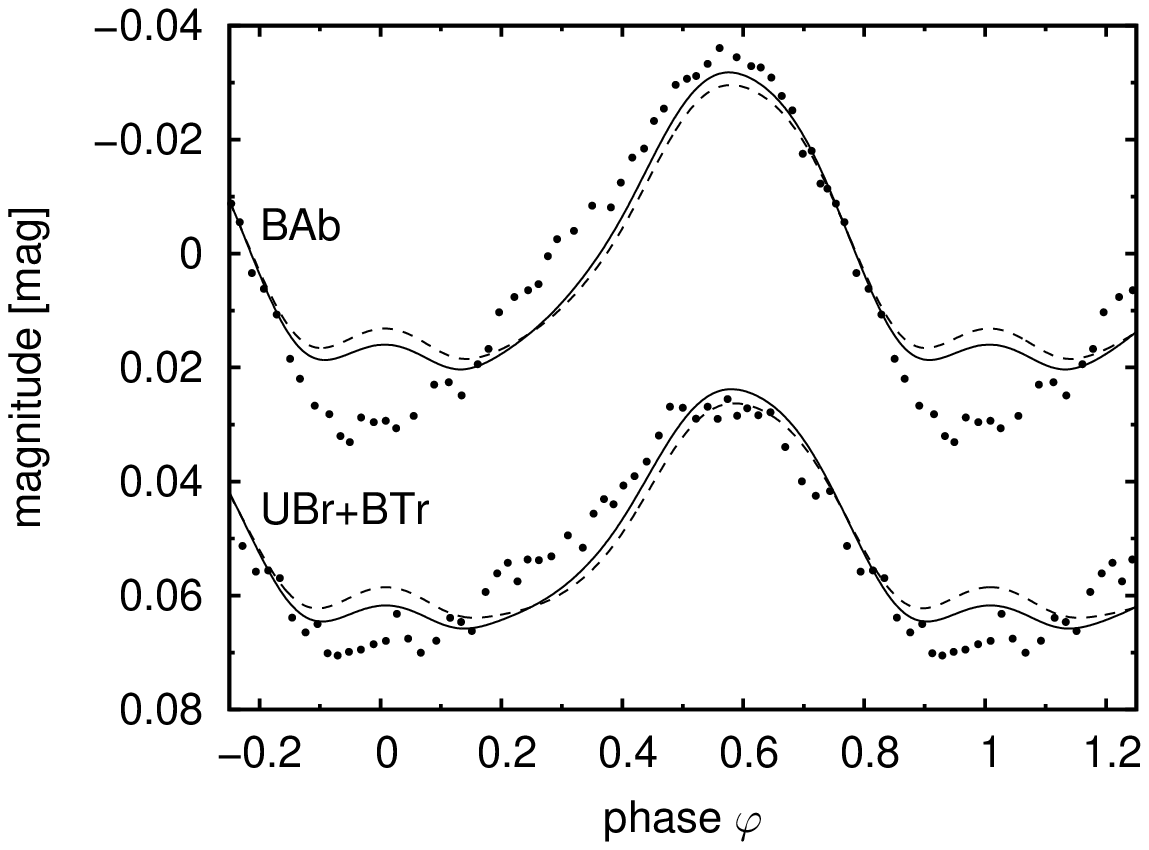}}
\caption{Same as Fig.~\ref{acen_hvvel}, however for BRITE observations in blue
(BRITE Austria, BAb) and red (UniBRITE, UBr, and BRITE Toronto, BTr) domains.
Data were averaged in phase intervals.}
\label{acen_hvvelbr}
\end{figure}

Simulated light curves were calculated from Eq.~\eqref{velik} as a function of
rotational phase. We used the abundance maps derived from MDI (\citealt{wadebrite},
Huang et al., in preparation) and the emergent fluxes computed with the SYNSPEC
code.

As a first step, we studied the influence of individual elements separately,
neglecting the distorted surface. We calculated visual light variations from the
abundance map of a single element, assuming fixed abundance of other elements
($\varepsilon_\text{He}=-1.0$, $\varepsilon_\text{N}=-3.9$,
$\varepsilon_\text{O}=-2.6$, $\varepsilon_\text{Si}=-4.5$, and
$\varepsilon_\text{Fe}=-4.4$). Fig.~\ref{prv_hvvel} shows that helium and iron
contribute principally to the visual light variations. These elements show large
overabundance in the spots and large abundance variations on the stellar
surface. The light curves reflect the abundance variations on the visible part
of the stellar surface. As a result of the anticorrelation of the helium and
iron surface distributions, the light curves due to these elements are also
anticorrelated. The influence of nitrogen, oxygen, and silicon on the visual
light variations is only marginal. This is not unexpected in the case of
nitrogen and oxygen, but the light curves due to the silicon show surprisingly
low amplitudes. The reason is the same for all these elements, because they do
not significantly influence the emergent flux (see Fig.~\ref{prvtoky}) and none
of these elements show large spots close to the center of the visible disk. A
small role of silicon in the overall light variability of \cpstar\ was already
unveiled by \citet{molacen}.

The light curve calculated from the surface distributions of all elements is
given in Fig.~\ref{acen_hvvel} for the Str\"omgren photometric system and in
Fig.~\ref{acen_hvvelbr} for the BRITE observations. The light curve is dominated
by iron in the optical region, and helium influences the light curve in the $u$
color. The adopted model explains most of the light variability in the optical
bands $vby$ of the Str\"omgren photometric system and in the blue and red bands
of BRITE. The inclusion of surface distortion via Eq.~\eqref{warpeq} dampens the
variations due to helium and leads to a better agreement between light curves
predicted for \cpstar\ and observed with the BRITE satellite
(Fig.~\ref{acen_hvvelbr}). For BRITE photometry, the inclusion of surface
distortion leads to a decrease of reduced $\chi^2$ from 19 to 13 in the blue
domain and from 10 to 7 in the red domain.

It remains unexplained why the shape of the light minimum during the phases
$\varphi\in[-0.1,0.4]$, where our model predicts local maximum due to helium, is
not reproduced by the observations. This is particularly clear in comparison
with the BRITE data. The local maximum due to helium is (in accordance with
Fig.~\ref{prvtoky}) strongest in the near-UV $u$ band. Helium dominates the
light curve due to large spot with $\varepsilon_\text{He}>0.5$ on the visible
hemisphere. The corresponding feature is completely missing in the observations.
As a result of this, the observed and predicted light curves disagree around
phase $\varphi\approx0.0$ in $u$. The disagreement in the $u$ colour could be
connected with the fact that the variations due to iron and helium partially
cancel in this domain. Consequently, the predicted light variations are
particularly sensitive to inaccuracies in the abundance maps or model opacities.
Omission of additional elements, NLTE effects or vertical abundance
stratification may also contribute to the disagreement between observation and
theory.  \citet{sokacen} concluded that besides iron also chromium and nickel
shape the UV flux distribution of \cpstar, thus possibly affecting the optical
light variability. \citet{wadebrite} discussed presence of pulsations with
similar amplitude as that due to distortion. However, their effect on the BRITE
curve is rather small, because period of pulsations and rotation are not
commensurable and the BRITE light curve was averaged in phase intervals.

\begin{table}
\caption{List of the IUE observations of \cpstar.}
\label{iuetab}
\centering
\begin{tabular}{ccccccc}
\hline
\multicolumn{3}{c}{SWP camera} & \multicolumn{3}{c}{LWR camera}\\
 Image & Julian date &  Phase & Image & Julian date &  Phase\\
        &   2,400,000+&&&2,400,000+\\
\hline
02112   &    43716.20883  & 0.036 &  01901 & 43716.20120  & 0.035 \\
02168   &    43722.19095  & 0.715 &  01941 & 43722.18401  & 0.714 \\
02175   &    73723.17280  & 0.826 &  01951 & 43723.16725  & 0.825 \\
02182   &    43724.14355  & 0.936 &  01965 & 43724.13660  & 0.935 \\
02199   &    43725.12749  & 0.048 &  01979 & 43725.12194  & 0.047 \\
02212   &    43726.17045  & 0.166 &  01990 & 43726.16490  & 0.165 \\
02223   &    43727.12593  & 0.274 &  01998 & 43727.11829  & 0.274 \\
02240   &    43729.12783  & 0.502 &  02039 & 43731.11237  & 0.727 \\
02257   &    43731.11793  & 0.727 &  02054 & 43732.77056  & 0.915 \\
02271   &    43732.77681  & 0.915 &  02089 & 43737.07990  & 0.403 \\
02287   &    43734.80857  & 0.146 &  02106 & 43739.13249  & 0.636 \\
02311   &    43737.08615  & 0.404  \\
02330   &    43739.14012  & 0.637  \\
\hline
\end{tabular}
\end{table}

\begin{figure*}
\centering \resizebox{0.49\hsize}{!}{\includegraphics{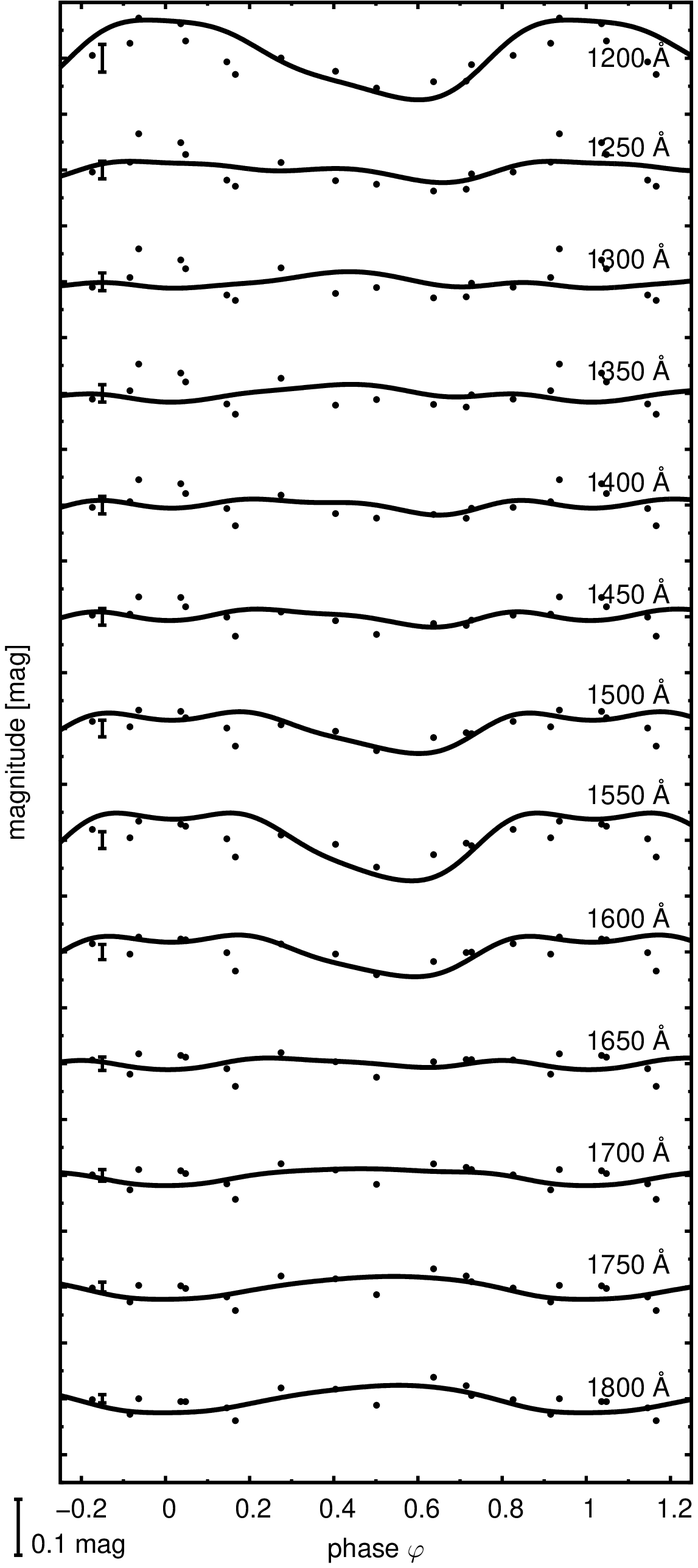}}
\centering \resizebox{0.49\hsize}{!}{\includegraphics{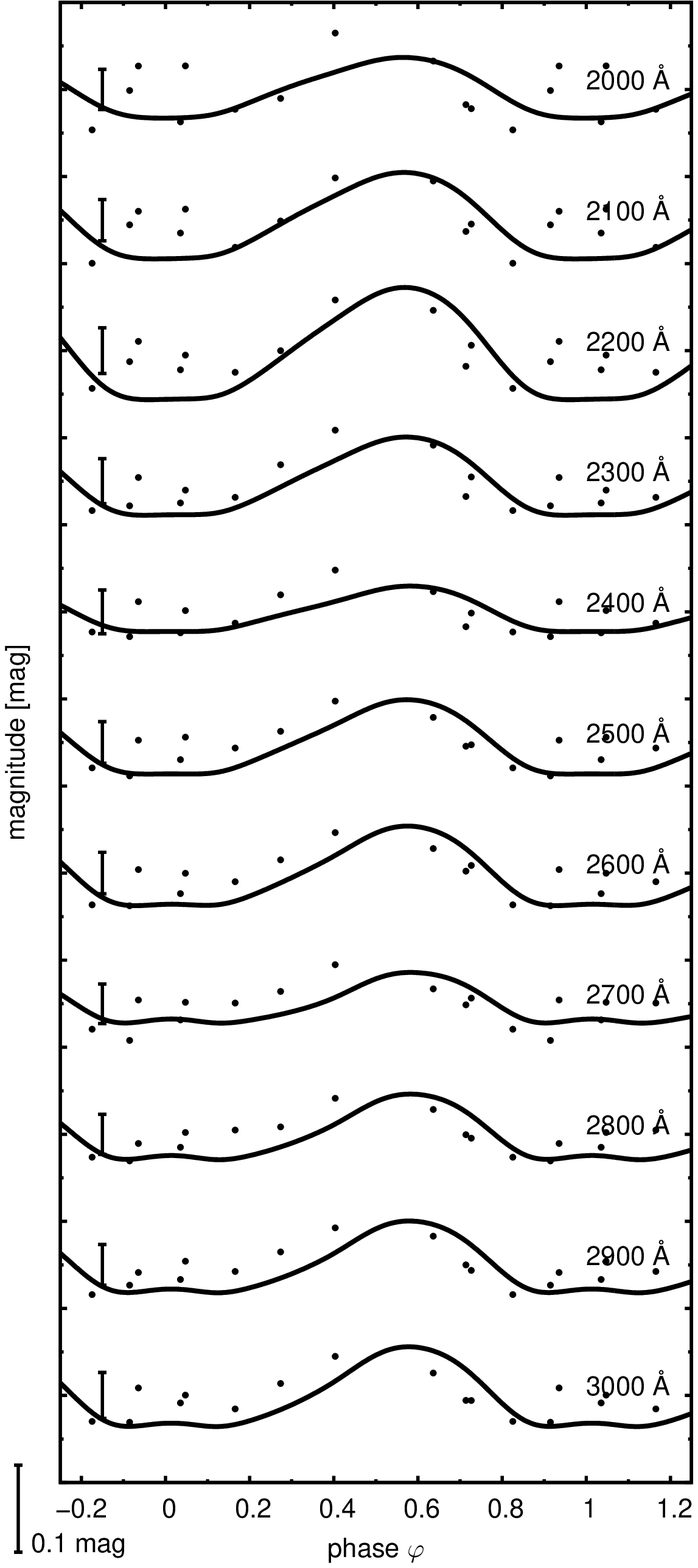}}
\caption{Comparison of the predicted (solid line) and observed (dots) UV light
variations for different wavelengths. Curves for individual wavelengths were
vertically shifted to more clearly illustrate the variability. The vertical
scale is given in the bottom left corner of each graph. Error bars, which are
plotted on the left side of the graphs, were derived from IUE observations at
each passband.}
\label{acenuv}
\end{figure*}

The light variability is caused by the flux redistribution from short wavelength
regions. This results in UV flux variations detected by \citet{molacen} and
\citet{sokacen}. To test the predicted UV light curves, we extracted
large-aperture, low-dispersion IUE spectra of \cpstar\ from the INES archive
\citep[see Table \ref{iuetab}]{ines} using the SPLAT package \citep{splat,pitr}.
The comparison of observed and predicted narrow-band UV variations calculated
using a Gaussian filter with dispersion of $50\,$\AA\ is given in
Fig.~\ref{acenuv}. Because the UV variations are dominated by the effect of
opacities and the effect of distortion is marginal in the UV domain, we plot
only the light curves that account for both surface opacity variations and
distortion. The comparison shows that the flux variations agree nicely, with the
difference between observations and simulation being comparable to the
observational uncertainty.

As follows from Fig.~\ref{prvtoky}, light variations in UV are dominated by
iron. Near-UV variations with $\lambda>1700\,$\AA\ are in phase with optical
variations \citep{sokacen}, because iron-rich regions are bright in this
wavelength domain (see Fig.~\ref{prvtoky}). On the other hand, iron line
absorption is especially important around $1550\,$\AA. For the shortest
wavelengths observed, around $1200\,$\AA, silicon and helium also influence the
flux variations. Our calculations agree with the results of \citet{sokacen}, who
also found that iron, silicon, and additionally carbon determine the flux
distribution and its variability in the UV spectral domain of \cpstar.

\begin{figure*}
\centering \resizebox{\hsize}{!}{\includegraphics{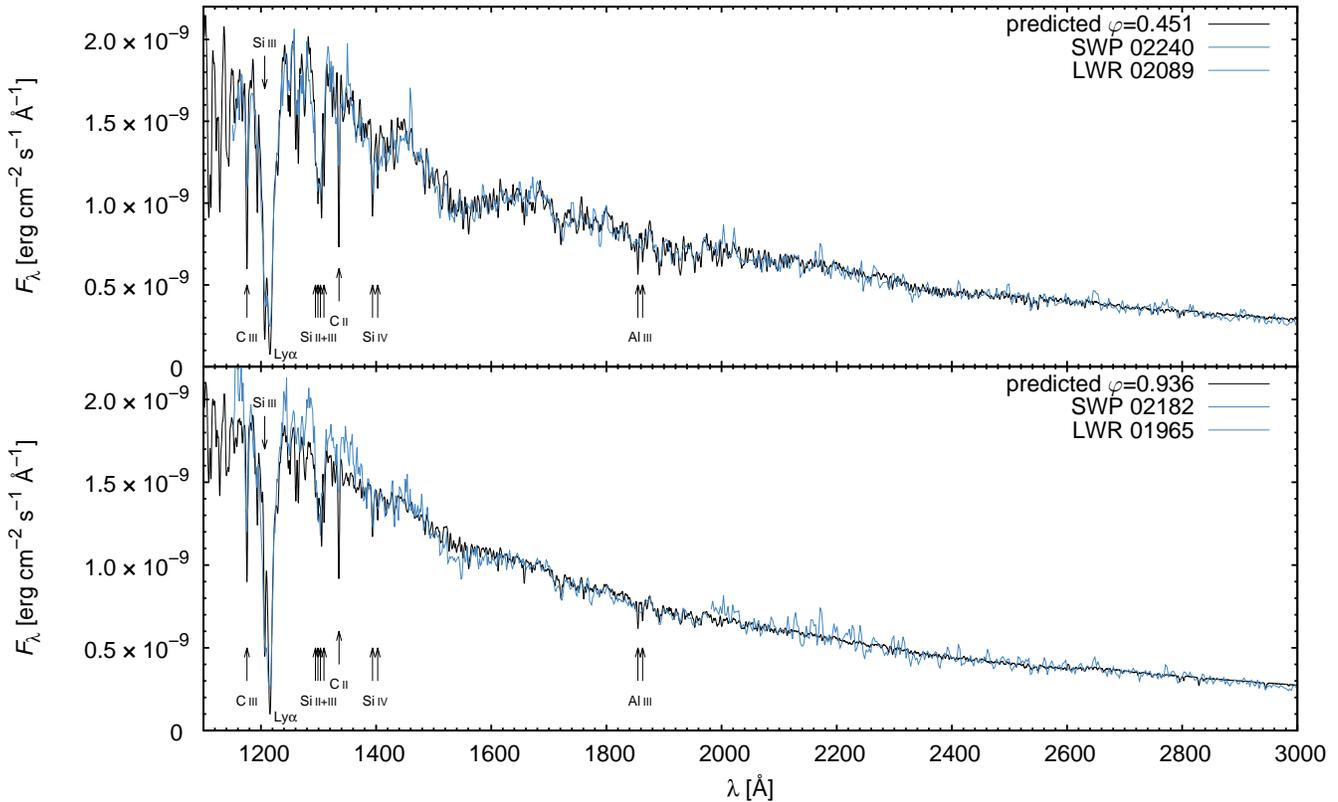}}
\caption{Comparison of the predicted (black line) and observed (blue line) UV
flux distribution for two rotational phases.}
\label{acenfm}
\end{figure*}

The simulated ultraviolet SED agrees nicely with the observed flux distribution at
individual rotational phases (Fig.~\ref{acenfm}). The flux distribution is
plotted for the rotational phases close to the maximum of iron and helium line
strength. 

\section{Conclusions}

We studied the effect of an aspherical distorted stellar surface on the
observational properties of helium chemically peculiar stars. Helium-rich
magnetic chemically peculiar stars show uneven surface distributions of helium.
As a result of this, the density scale height varies across the stellar surface.
Consequently, helium-rich regions are depressed inwards and show effectively
lower local radius and higher effective temperature, keeping the total
luminosity constant. From the photometric point of view, helium-rich regions
become slightly brighter in the far-ultraviolet region due to the distortion of
the stellar surface, while becoming fainter in the near-ultraviolet and optical
regions.

We demonstrated the effect of the distorted surface on helium-rich magnetic
chemically peculiar star a~Cen. We calculated model atmospheres with a chemical
composition corresponding to the abundance maps derived using magnetic Doppler
imaging of this star. We integrated the emergent flux across the chemically
inhomogeneous stellar surface to predict the light variability. We have shown
that most of the light variability is due to the redistribution of the flux from
far-ultraviolet to the near-ultraviolet and optical regions due to variable
blanketing caused by iron line transitions and helium bound-free transitions. 

The effect of surface distortion causes additional light variability with
amplitudes of the order of millimagnitudes. It contributes to the light
variability caused by variable blanketing and leads to improvement of the
agreement between predicted and observed light curves in the optical region.
However, the modification of the light curve caused by the surface distortion is
smaller than the difference between predicted and observed light curves.
Consequently, additional effects likely contribute to the observed light curve.
The effect of distortion is negligible in the UV region, where the amplitude of
the light variability is larger, and where the flux redistribution due to the
abundance spots nicely reproduces the observed flux variations.

Although surface distortion is not a dominant effect for the light variability
of chemically peculiar stars, it affects photometry with millimagnitude
precision. Therefore, it is required to obtain good detailed agreement between
simulated and observed light curves of helium rich chemically peculiar stars.

\section*{Acknowledgements}

The authors thank Drs.~L.~Huang and J.~Silvester for providing us with the
surface abundance maps. This work was supported by grant GA \v{C}R 18-05665S.
GAW acknowledges Discovery Grant support from the Natural Science and
Engineering Research Council (NSERC) of Canada. Based on data collected by the
BRITE Constellation satellite mission, designed, built, launched, operated and
supported by the Austrian Research Promotion Agency (FFG), the University of
Vienna, the Technical University of Graz, the University of Innsbruck, the
Canadian Space Agency (CSA), the University of Toronto Institute for Aerospace
Studies (UTIAS), the Foundation for Polish Science \& Technology (FNiTP MNiSW),
and National Science Centre (NCN). Access to computing and storage facilities
owned by parties and projects contributing to the National Grid Infrastructure
MetaCentrum provided under the program "Projects of Large Research, Development,
and Innovations Infrastructures" (CESNET LM2015042) is greatly appreciated.

\bsp

\label{lastpage}

\end{document}